\documentclass[aps,prb,reprint,showpacs,superscriptaddress,longbibliography]{revtex4-1}
\usepackage{graphicx}
\usepackage{amsmath,amssymb,amsfonts}
\usepackage{comment}
\usepackage{color}
\usepackage[T1]{fontenc}
\usepackage{textcomp}
\usepackage{hyperref}
\hypersetup{
setpagesize=false,
colorlinks=true,linkcolor=blue,
citecolor=red,
}
\usepackage{bm,ulem} 
\usepackage{bm}
\usepackage{mathrsfs}
\usepackage{physics}
\usepackage{ulem}
\usepackage{braket}
\usepackage[version=4]{mhchem}
\makeatletter
\let\MYcaption\@makecaption
\makeatother
\usepackage{subcaption}
\makeatletter
\let\@makecaption\MYcaption
\makeatother

\begin{document}
\title{Spin supercurrent 
in parity mixed superconductors with structural chirality}
\author{Keito Hara}
\email{hara.keito.43e@st.kyoto-u.ac.jp}
\affiliation{%
Department of Physics, Kyoto University, Kyoto 606-8502, Japan
}%
\author{Youichi Yanase}
\affiliation{%
Department of Physics, Kyoto University, Kyoto 606-8502, Japan
}%

\begin{abstract}
 Chiral materials exhibit a spin filtering effect, so-called chirality-induced spin selectivity (CISS). A recent observation of spin accumulation at the ends of a chiral-structured superconductor has opened up a new pathway for studying the CISS effect in superconductors. In chiral-structured superconductors, the admixture of the spin-singlet and spin-triplet order parameters significantly influences the properties of superconductivity. In this paper, we investigate the interplay between the superconducting order parameter and supercurrent-induced spin current, namely, the superconducting CISS effect. In weakly party-mixed superconductors,
 the spin current, which is predominantly temperature-independent, is carried by spin-polarized Cooper pairs with finite center-of-mass momentum. In contrast, in strongly party-mixed superconductors
 the temperature-dependent spin current is also carried by electrons with opposite momentum and antiparallel spins forming a Cooper pair. Chiral-structured superconductors will offer a novel platform for exploring the CISS effect and may provide deeper insights into its underlying mechanisms related to the parity-mixed order parameter.
\end{abstract}
\maketitle

\section{Introduction}
Chirality represents a structure that lacks both inversion and mirror symmetries. When electrons pass through chiral materials, their spins become polarized parallel or antiparallel to the direction of the current, a phenomenon known as the chirality-induced spin selectivity (CISS) effect~\cite{Ray1999cv,Gohler2011-zo,Naaman2012-pa,Naaman2015-rb,Naaman2019-nl,Naaman2020-mr,Waldeck-APL2021,chem-2024-Bllom}. The CISS effect, initially identified in organic molecules, has now been demonstrated in inorganic materials~\cite{ chem-ghosh-2019, Inui2020-yl, prl-siota-2021, nano-2022-mollers, Suzuki2023-or, Kato2024-uk}. Despite its broader experimental platform~\cite{Qian2022-bp}, the underlying mechanisms of the CISS effect remain elusive. A recent experiment has demonstrated the accumulation of antiparallel spins at the ends of a chiral-structured superconductor under alternating electric currents~\cite{Nakajima2023-ge}. Owing to the long-range coherence of the superconducting wave function, the spin currents generated by the electric current can propagate over a macroscopic distance~\cite{Nakajima2023-ge}. This observation highlights the potential use of superconductors with structural chirality~\cite{Togano2004-re,Badiba-jpsj2005,Lee-prb-2005,Yuan2006-lg,harada-prb-2012,maruyama-2013-jpsj,Alpern2016-fp,Qin2017-NatCom,Shapira2018-dt,Carnicom2018-tj,Alpern2019-mw,Lv-prb-2020,Alpern2021-cb,He-naturecom-2022,Nakamura-jpsj-2023,Gao2023-prb,Wan2023-hn,Nakajima2024-bz,Afzal2024-al,Watanabe2024} as a promising platform for the study of the CISS effect~\cite{Yao2024-qj,sato2025}.

The inversion symmetry breaking in superconductors with structural chirality has a profound influence on the superconducting order parameter~\cite{Bauer2012-ix,Smidman_2017}. In materials without an inversion center, spin and parity can no longer independently classify Cooper pairs, leading to the admixture of even-parity spin-singlet and odd-parity spin-triplet pairing. Then, unlike conventional $s$-wave superconductors, the superconducting gap becomes anisotropic 
and may exhibit accidental nodes on the Fermi surface, which have a significant impact on the properties of superconductivity~\cite{Gorkov2001-uf,Togano2004-re,Yogi2004-cy,Badiba-jpsj2005,Lee-prb-2005,Sato2006-is,Yuan2006-lg,Hayashi2006-vf,Hayashi2006-ze,SIGRIST2007536,fujimoto-jpsj-2007,Yanase2008-ra,Samokhin-prb-2008,harada-prb-2012,maruyama-2013-jpsj}. The purpose of our study is to elucidate the interplay between the CISS effect and the superconducting order parameter in chiral-structured superconductors.

In this paper, we investigate chiral-structured superconductors with various order parameters. To discuss spin accumulation at the ends of superconductors, we calculate the longitudinal spin current parallel to the electric current. While the electric current is odd under time-reversal operation, the spin current is even. Therefore, the electric current should induce a nonlinear spin current proportional to the square of the electric current~\cite{Hamamoto2017-kr}, although the spin current linear to the electric current is prohibited. This constraint aligns with the experimental observation of antiparallel spin accumulation at the ends of chiral-structured superconductors under an alternating electric current~\cite{Nakajima2023-ge}. The calculated results show that, in superconductors without the accidental nodal structure of the superconducting gap, the temperature-independent spin current is carried primarily by spin-polarized Cooper pairs with finite center-of-mass momentum. On the other hand, when the superconducting gap is sufficiently small or vanishes due to the parity mixing in the order parameter, the temperature-dependent spin current is also carried by electrons with opposite momentum and antiparallel spins forming a Cooper pair. Furthermore, in strongly parity-mixed superconductors, the spin current can be enhanced at low temperatures far below the transition temperature, primarily due to quasiparticle excitations. The mechanism underlying the CISS effect remains a long-standing mystery, and our study of the superconducting CISS effect may offer critical insights into its origin.

\section{Results}
\subsection{Model}
\begin{figure}
	\includegraphics[width=3.0in]{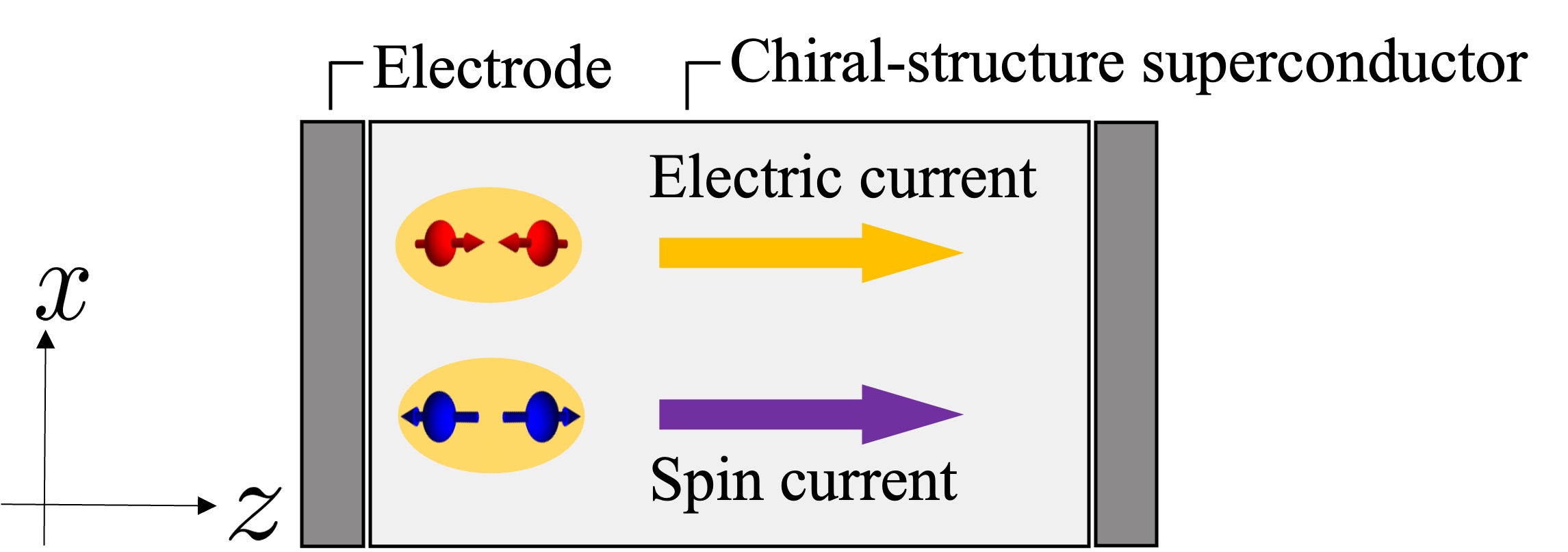}
	\caption{Illustration of the supercurrent-induced spin current. A superconductor with structural chirality hosts a spin current under an electric current. The spin current causes anti-parallel spin accumulation at the edges of the superconductor.}
	\label{setup}
\end{figure}

A setup of the system is schematically shown in Fig.~\ref{setup}. We consider a two-dimensional superconductor with structural chirality under an applied electric current. Lack of inversion symmetry gives rise to antisymmetric spin-orbit coupling (ASOC), which generates a spin current driven by the electric current. To describe this phenomenon, we introduce a mean-field Hamiltonian
\begin{equation} 
\begin{split}
H_{\rm MF}^{\vb*{q}}
&= \sum_{\vb*{k}ss'}
\left[(\epsilon(\vb*{k})-\mu){\delta}_{ss'} +\alpha \vb*{g}(\vb*{k})\vdot\vb*{\sigma}_{ss'} \right]
c_{\vb*{k}s}^\dag c_{\vb*{k}s'}\\
&+\frac{1}{2}\sum_{\vb*{k}ss'}
\Bigl(
c_{\vb*{k}+\frac{\vb*{q}}{2}s}^\dag 
c_{-\vb*{k}+\frac{\vb*{q}}{2}s'}^\dag
\Delta_{ss'}(\vb*{k})
+ {\rm H.c}\Bigr),
\end{split}
\end{equation} 
where $\epsilon(\vb*{k}) = -2t (\cos k_x + \cos k_z)$ and $\vb*{g}(\vb*{k}) = (\sin k_x,\ 0,\ \sin k_z)$ represent the hopping energy and the ASOC, respectively. Here, $t$ denotes a hopping parameter in the tight-binding model, $\mu$ represents the chemical potential, $s$ is an index of spin, and the lattice constant is set to unity. Without loss of generality, we assume $\alpha >0$. The parameters are given by $(t,\ \mu,\ \alpha) = (1.0,\ -1.5,\ 0.6)$ unless mentioned otherwise. In noncentrosymmetric superconductors, even-parity spin-singlet pairing and odd-parity spin-triplet pairing are generally admixed~\cite{Bauer2012-ix}. Consequently, the superconducting gap function is given in a parity-mixed form $\Delta_{ss'}(\vb*{k}) = \mqty[i\gamma(\vb*{k})(\psi_{\rm s}
+ \psi_{\rm t}\vb*{g}(\vb*{k})\vdot \vb*{\sigma})\sigma_y]_{ss'}$, where $\gamma(\vb*{k})$ is an even function that transforms according to an irreducible representation of the point group of the system $C_4$. In this paper, we study superconductivity with $s$-wave, $p$-wave, and $s$$+$$p$-wave symmetry, specifically for $\gamma(\vb*{k}) = 1$, and $d$-wave symmetry for $\gamma(\vb*{k}) = \cos k_x -\cos k_z$. Note that the superconducting gap exhibits nodes on the Fermi surfaces when $\gamma(\vb*{k}) = \cos k_x -\cos k_z$. 

In this setup, the non-dissipative electric current is carried by the Cooper pair with finite momentum $\vb*{q}$. We consider an electric current flowing along the $z$ direction, and the center-of-mass momentum of Cooper pairs is restricted to $\vb*{q} = q\hat{z}$. For a given $\vb*{q}$ and a temperature $T$, the order parameters $\psi_s$ and $\psi_t$ are determined by the gap equation, and thus depend on $\vb*{q}$. However, at sufficiently low temperatures, the dependence of $\psi_s$ and $\psi_t$ on $\vb*{q}$ becomes almost negligible. In the following, we set the parameters $\psi_s,\ \psi_t >0$ and $\psi = \psi_s +\psi_t = 0.3$, where $\psi$ serves as an indicator of the superconducting transition temperature $T_{\rm c}$ and the maximal superconducting gap.

We show quasiparticle energy bands in the absence of electric current. The Hamiltonian in normal state is written as
$H_{\rm N}(\vb*{k})=\epsilon(\vb*{k})-\mu +\alpha \vb*{g}(\vb*{k})\vdot\vb*{\sigma}$, and
the energy dispersion of the spin-splitting bands is given by $\xi_\pm (\vb*{k}) = \epsilon(\vb*{k}) - \mu \pm \alpha |\vb*{g}(\vb*{k})|$. Note that the spin and $\vb*{g}$-vector, $\vb*{g}(\vb*{k})$, are parallel or antiparallel corresponding to the index $\pm$ of the energy bands, namely the $+$ and $-$ bands, and the spin is opposite between $\vb*{k}$ and $\vb*{-k}$. Fermi surfaces are defined by $\xi_\pm(\vb*{k}) = 0$. Energy dispersion of Bogoliubov quasiparticles is given by 
\begin{equation} 
E_\pm(\vb*{k}) = \sqrt{\xi_\pm^2(\vb*{k})+\Delta_\pm^2(\vb*{k})},
\end{equation} 
since the interband pairing is absent in our model.
Here, $\Delta_\pm(\vb*{k}) = \gamma(\vb*{k})(\psi_s \pm\psi_t|\vb*{g}(\vb*{k})|)$ is regarded as the gap function of the $\pm$ bands. Thus, the intraband Cooper pair of antiparallel spins with zero center-of-mass momentum is formed in the absence of the electric current.

\begin{figure}
	\includegraphics[width=3.4in]{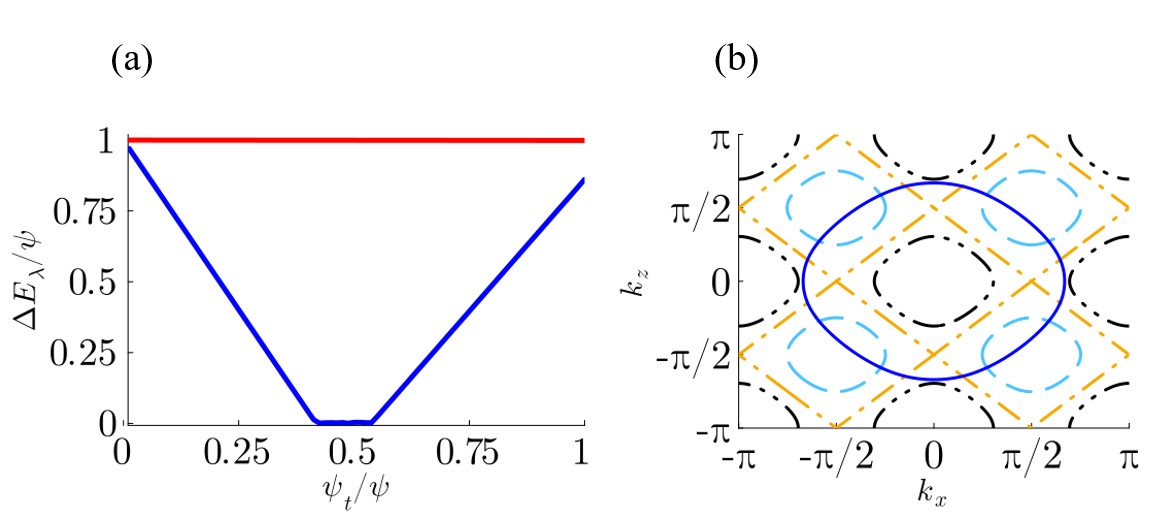}
	\caption{(a) The order parameter dependence of the excitation gap $\Delta E_\lambda = \min_{\vb*{k}}E_\lambda(\vb*{k})$ normalized by $\psi$ for $\lambda = +$ (red line) and $-$ (blue line). (b) The Fermi surface of the $-$ band (blue line) and the zeros of the gap function $\Delta_{-}(\vb*{k})$ for $\psi_t/\psi = 0.45$ (cyan dashed line), $0.5$ (yellow dashed line) and $0.55$ (black dashed line).}
	\label{gap}
\end{figure}

Different from isotropic $s$-wave superconductors, the order parameters in the band basis $\Delta_\pm(\vb*{k})$ may have some zeros due to the factor $\psi_s \pm\psi_t|\vb*{g}(\vb*{k})|$. Thus, a gap node may appear on Fermi surfaces. The order parameter dependence of the excitation gap in the $s$$+$$p$-wave superconducting state is shown in Fig.~\ref{gap}(a). 
The excitation gap vanishes in the $-$ band when the parity mixing is significant $\psi_t/\psi \simeq 0.5$.
The Fermi surface $\xi_-(\vb*{k})=0$ and the zeros of the gap function $\Delta_-(\vb*{k}) = 0$ are plotted in Fig.~\ref{gap}(b), revealing that energy spectrum in the $-$ band has nodes and the electron-like Bogoliubov band $E_-(\vb*{k})$ touches the hole-like band $-E_-(\vb*{k})$. 

\begin{figure}
\includegraphics[width=2.6in]{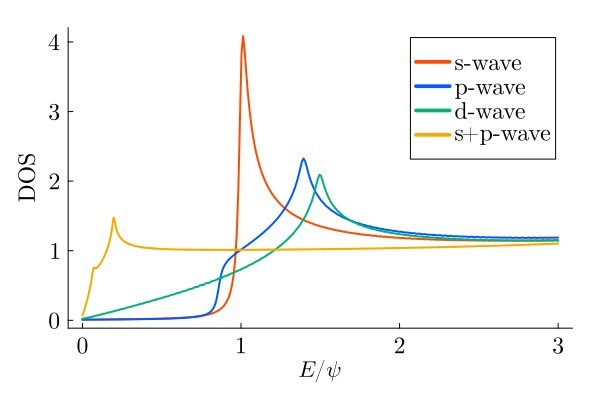}
\caption{Calculated DOS of the $-$ band in the $s$-wave, $p$-wave, $d$-wave, and $s$$+$$p$-wave ($\psi_t/\psi = 0.5$) superconducting states. The DOS is normalized as $\sum_{\vb*{k}}\delta(E-E_-(\vb*{k}))/\sum_{\vb*{k}}\delta(\xi_-(\vb*{k}))$.}
\label{DOS}
\end{figure}

The density of states (DOS) of the $-$ band for various order parameters is shown in Fig.~\ref{DOS}. The result reveals that the DOS in the $s$-wave and $p$-wave superconducting states vanishes for $E/\psi \lesssim 1$ while the DOS is finite in the $s$$+$$p$-wave state with $\psi_t/\psi = 0.5$ and the $d$-wave state due to the nodal gap structure. Furthermore, the low-energy DOS at $E/\psi \lesssim 1$ in the nodal $s$$+$$p$-wave superconductors ($\psi_t/\psi = 0.5$) is much larger than that in the $d$-wave superconductors. As we show later, these features in the quasiparticle excitations are responsible for the mechanism and temperature dependence of the CISS effect. 

\subsection{Symmetry analysis}
To estimate the superconducting CISS effect, we calculate the spin current induced by the supercurrent. We assume time-reversal symmetry in the absence of the electric current. Note that the effective time-reversal symmetry breaking due to dissipation does not occur in the non-dissipative transport of superconductors. The electric current $j_{z}$ and the center-of-mass momentum $q$ are odd under a time-reversal operation, whereas the spin current $j_{z}^{s_z}$ is even. Thus, we obtain the expansion
\begin{equation} 
j_{z}(q) = D_z q + \mathcal{O}(q^3),
\end{equation} 
\begin{equation} 
j_{z}^{s_z}(q) = {\rm const.} + D_{z}^{s_z} q^2 + \mathcal{O}(q^4),
\end{equation}
where $D_z$ is the superfluid density~\cite{tinkam}.  The coefficient $D_z^{s_z}$ can be regarded as a nonreciprocal spin superfluid density analogous to the nonreciprocal superfluid density~\cite{watanabe-prb-2022,watanabe-prb-2022-2}. We find that the electric current induces the nonreciprocal spin current proportional to the square of the electric current as
\begin{equation} 
\label{gamma}
j_{z}^{s_z} = {\rm const.} + \gamma j_z^2 + \mathcal{O}(j_z^4),\quad \gamma = \frac{D_{z}^{s_z}}{(D_z)^{2}}.
\end{equation}
The supercurrent-induced spin current is characterized by the coefficient $\gamma$.
In this paper, we calculate $D_z$ and $D_z^{s_z}$ for various order parameters to clarify the interplay between the CISS effect and the pairing states. 

A similar symmetry argument is applied when the system has either inversion or mirror symmetry. Because both the spin current $j_{z}^{s_z}$ and the center-of-mass momentum $q$ are odd under inversion and mirror symmetry flipping $q$, the spin current is an odd function of the center-of-mass momentum, which contradicts the argument based on time-reversal symmetry. Thus, the symmetry analyses indicate that supercurrent-induced spin current requires the breaking of both inversion and mirror symmetry, namely structural chirality. In our model, the ASOC and the parity mixing in superconducting order parameters break these symmetries.

\subsection{Electric current}
\begin{figure}
\includegraphics[width=3.2in]{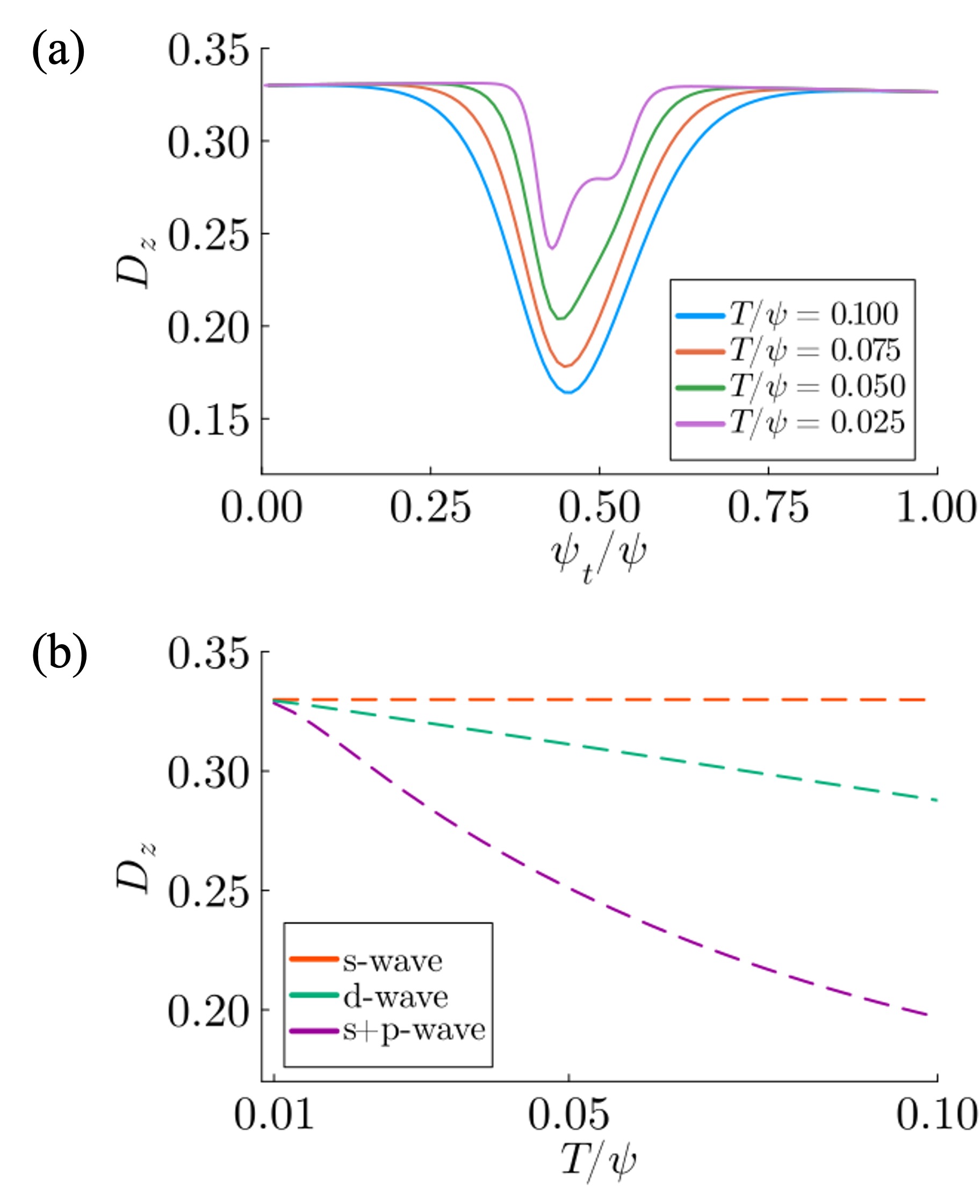}
\caption{(a) Order parameter dependence of the superfluid density $D_z$ in the $s$$+$$p$-wave superconducting state at several temperatures. (b) Temperature dependence of the superfluid density in the $s$-wave, $d$-wave, and $s$$+$$p$-wave ($\psi_t/\psi = 0.4$) superconducting states.}
\label{Dz_T}
\end{figure}

First, we show the results for the electric current. The supercurrent generated by the center-of-mass momentum $q$ is characterized by the superfluid density $D_z$, whose order parameter dependence in the $s$$+$$p$-wave superconductors is shown in Fig.~\ref{Dz_T}(a). It is shown that the parity-mixing suppresses the superfluid density. Furthermore, in strongly parity-mixed superconductors ($\psi_t/\psi \simeq 0.5$) which leads to the nodal superconducting gap, the superfluid density decreases with increasing temperature. In contrast, in weakly parity-mixed superconductors ($\psi_t/\psi \simeq 0,1$), the superfluid density is nearly temperature independent, assuming that the order parameters are almost temperature independent in this range. Therefore, the suppression in the superfluid density is attributed to excited quasiparticles.

The temperature dependence of the superfluid density is shown in Fig.~\ref{Dz_T}(b), demonstrating that the suppression of the electric current by elevating temperature is more pronounced in the $s$$+$$p$-wave state than in the $d$-wave state. This is consistent with the fact that the low-energy quasiparticle DOS of $s$$+$$p$-wave superconductors is notably higher than that of $d$-wave superconductors (Fig.~\ref{DOS}). The $s$$+$$p$-wave superconducting state suffers from the effects of quasiparticle excitations more significantly than the $d$-wave one.

To elucidate the physical meaning behind the suppression of the electric current, we decompose the electric current into two components as follows (see the Methods section for derivation).
\begin{equation} 
j_z(q)
= j_z^{(1)}(q) + j_z^{(2)}(q),
\end{equation} 
\begin{equation} 
\begin{split}
j_z^{(1)}(q)
&= \frac{2}{V}\sum_{\vb*{k}}v_{k_z}^{(0)}n_{\vb*{k}}^{(1)}q + \mathcal{O}(q^2)\\
&= D_z^{(1)} q + \mathcal{O}(q^2),
\end{split}
\end{equation} 
\begin{equation} 
\begin{split}
\label{dia}
j_z^{(2)}(q)
&= \frac{2}{V}\sum_{\vb*{k}}v_{k_z}^{(1)}n_{\vb*{k}}^{(0)}q + \mathcal{O}(q^2)\\
&= D_z^{(2)} q + \mathcal{O}(q^2),
\end{split}
\end{equation} 
where the electron charge is set as $e=1$ and $V$ is the system size. The electric current $j_z^{(1)}$ and $j_z^{(2)}$ are called the paramagnetic current and the diamagnetic current, respectively~\cite{tinkam}. Here, $v_{k_z}^{(i)}$ and $n_{\vb*{k}}^{(i)}$ are defined as
\begin{equation} 
v_{k_z+q/2\uparrow} + v_{-k_z+q/2\downarrow}
= 2 v_{k_z}^{(1)}q+ \mathcal{O}(q^3),
\end{equation} 
\begin{equation} 
v_{k_z+q/2\uparrow} - v_{-k_z+q/2\downarrow}
= 2(v_{k_z}^{(0)} + v_{k_z}^{(2)}q^2) + \mathcal{O}(q^4),
\end{equation} 
\begin{equation} 
\braket{n_{\vb*{k}+q/2\uparrow}}_q
+ \braket{n_{-\vb*{k}+q/2\downarrow}}_q 
= 2(n_{\vb*{k}}^{(0)} +  n_{\vb*{k}}^{(2)} q^2 ) 
+ \mathcal{O}(q^4),
\end{equation} 
\begin{equation} 
\braket{n_{\vb*{k}+q/2\uparrow}}_q
- \braket{n_{-\vb*{k}+q/2\downarrow}}_q
= 2n_{\vb*{k}}^{(1)} q + \mathcal{O}(q^3),
\end{equation}
with the velocity $v_{k_zs}=\partial_{k_z}H_{\rm N}(\vb*{k})_{ss}$ and the particle number operator $n_{\vb*{k}s}=c_{\vb*{k}s}^\dag c_{\vb*{k}s}$. The statistical average $\braket{\cdots}_q = {{\rm tr}[\cdots e^{-H_{\rm MF}^q/T}]}/{{\rm tr}[e^{-H_{\rm MF}^q/T}]}$ is obtained with the mean-field Hamiltonian $H_{\rm{MF}}^{q}$. These equations indicate that $2(v_{k_z}^{(0)}+v_{k_z}^{(2)}q^2)$ and $v_{k_z}^{(1)}q$ represent the relative velocity and the center-of-mass velocity of electrons forming Cooper pairs, respectively. In the same way, $2n_{\vb*{k}}^{(0)}$ is regarded as the density of Cooper pairs in the absence of electric current, while $2n_{\vb*{k}}^{(1)}q$ represents current-induced spin magnetization, a phenomenon known as the Edelstein effect~\cite{Edelstein1990-cb,Edelstein1995-fq,He-prr-2020,Ikeda2020-lo}. Therefore, Eq.~(\ref{dia}) shows that the diamagnetic current $j_z^{(2)}$ is carried by Cooper pairs with center-of-mass velocity $v_{k_z}^{(1)}q$ and density $2n_{\vb*{k}}^{(0)}$. 

\begin{figure}
\includegraphics[width=2.8in]{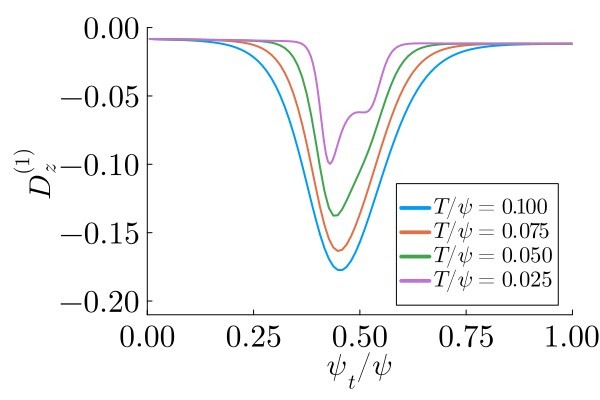}
\caption{The paramagnetic component of the superfluid density $D_z^{(1)}$ in the $s$$+$$p$-wave superconducting state at various temperatures.}
\label{Dz_1}
\end{figure}

The diamagnetic current is partly canceled by the paramagnetic current $j_z^{(1)}$ proportional to $D_z^{(1)}$,
which is plotted in Fig.~\ref{Dz_1}. The paramagnetic current is largely negative in the nodal $s+p$-wave state and shows order parameter dependence similar to the electric current $j_z$. Consequently, the decrease in the electric current $j_z$ shown in Fig.~\ref{Dz_T}(a) is primarily attributed to the paramagnetic current $j_z^{(1)}$ carried by quasiparticles, consistent with the previous discussion. 

\begin{figure}[hb]
\includegraphics[width=2.6in]{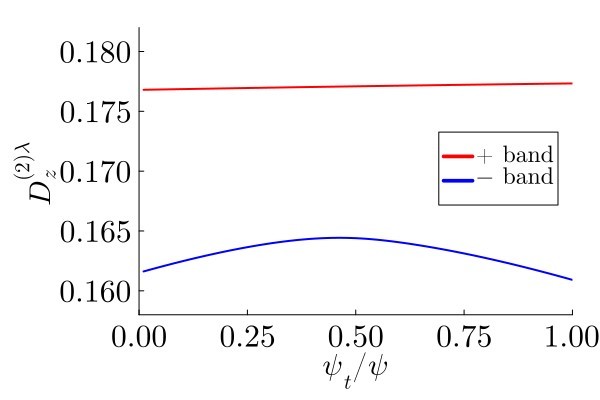}
\caption{The diamagnetic component of the superfluid density arising from the $\lambda(=\pm)$ band $D_z^{(2)\lambda}$ in the $s$$+$$p$-wave superconducting state at $T/\psi = 0.1$.}
\label{Dz2}
\end{figure}

\begin{figure*}[ht]
\begin{minipage}{0.45\linewidth}
\hspace*{-0.5cm}
\includegraphics[width=2.8in]{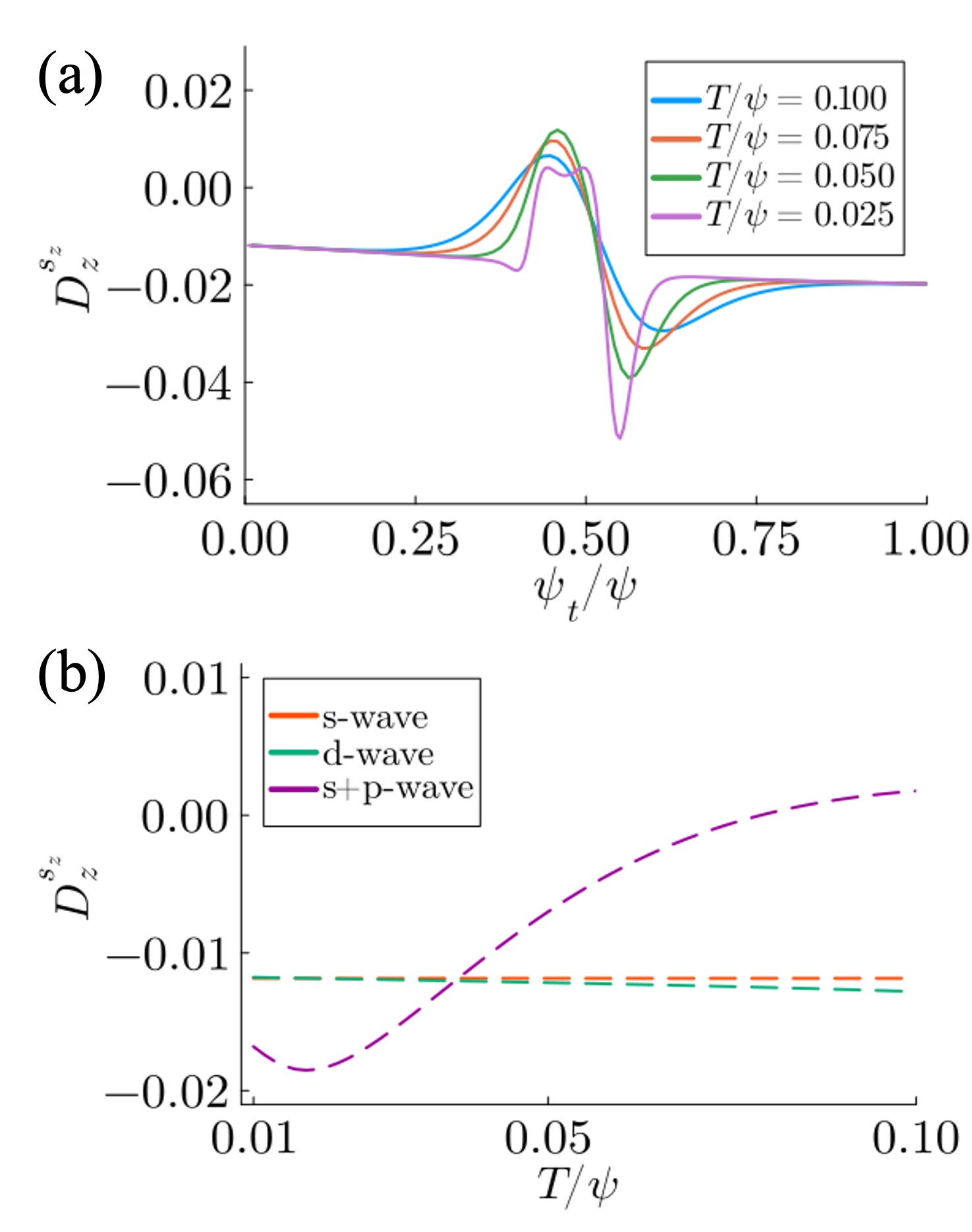}
\end{minipage}
\begin{minipage}{0.45\linewidth}
\centering
\includegraphics[width=2.8in]{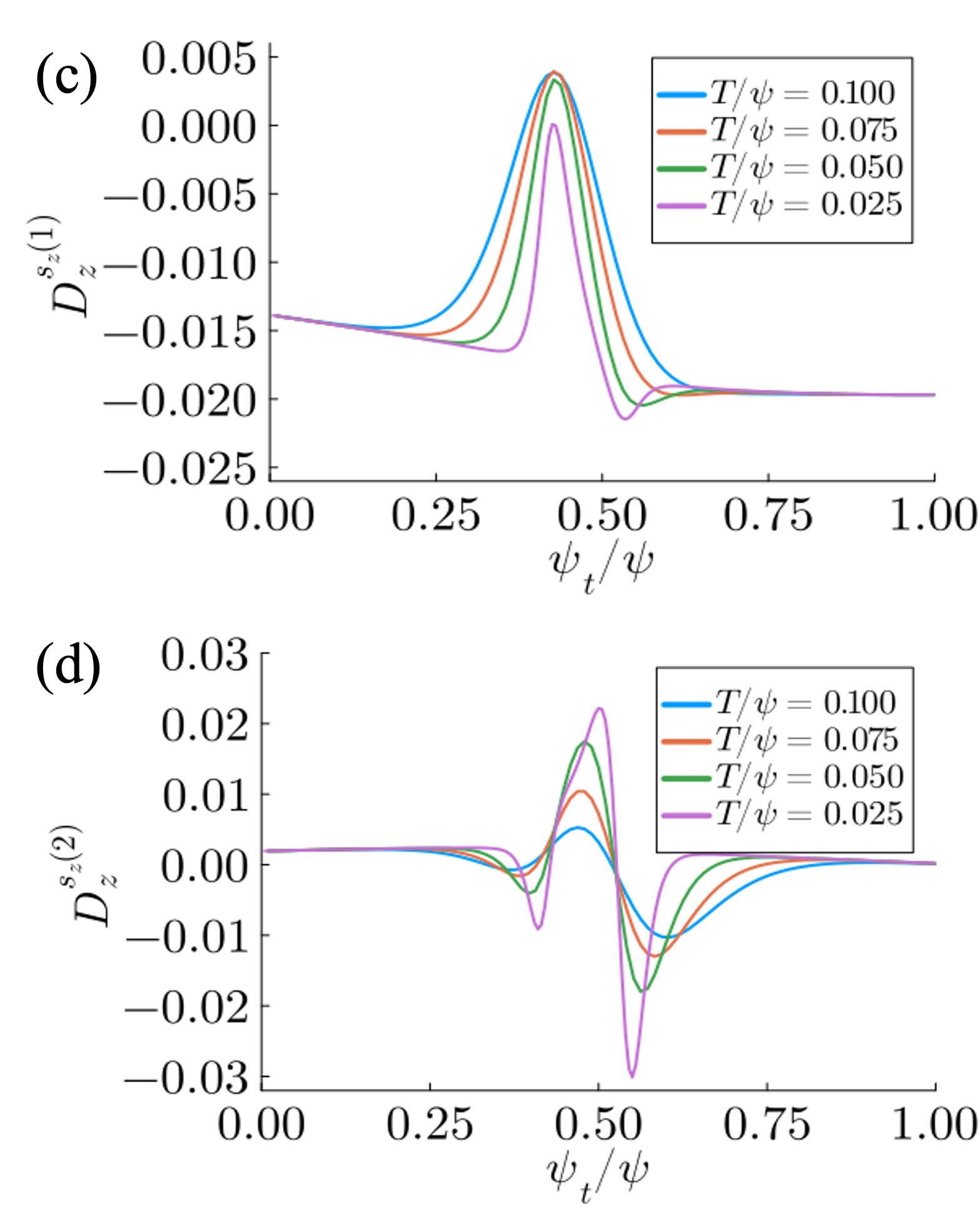}
\end{minipage}
\caption{(a) Order parameter dependence of the spin superfluid density $D_z^{s_z}$ in the $s$$+$$p$-wave superconducting state at various temperatures. (b) Temperature dependence of the spin superfluid density in the $s$-wave, $d$-wave, and $s$$+$$p$-wave ($\psi_t/\psi = 0.4$) superconducting states. (c) $D_z^{s_z(1)}$ and (d) $D_z^{s_z(2)}$ in the $s$$+$$p$-wave superconducting state.} 
\label{Dzsz} 
\end{figure*}

The diamagnetic current is proportional to $D_z^{(2)}$, which is written as
\begin{equation}
\begin{split}
\label{eq:Dz2}
D_z^{(2)} =& D_z^{(2)+} + D_z^{(2)-},\\
D_z^{(2)\lambda} =& \frac{1}{V}\sum_{\vb*{k}} (2t \cos k_z - \alpha \sin k_z) \\
& \times \cos \alpha_{\vb*{k}}^\lambda  
\frac{1 + \lambda \cos\theta_{\vb*{k}}}{2}
\qty(f(E_\lambda(\vb*{k}))- \frac{1}{2}),
 \end{split}
 \end{equation} 
with the Fermi distribution function $f(x)$. Here, we have defined
 \begin{equation}
 \begin{split}
 \label{alpha_def}
\cos \theta_{\vb*{k}}
&= \frac{ g_z(\vb*{k})}{| \vb*{g}(\vb*{k})|},
\
\sin \theta_{\vb*{k}}
= \frac{ |g_x(\vb*{k})|}{| \vb*{g}(\vb*{k})|},\\
\cos \alpha_{\vb*{k}}^\lambda 
&=
\frac{
\xi_\lambda(\vb*{k})
}{E_\lambda(\vb*{k})},
\
\sin \alpha_{\vb*{k}}^\lambda 
= 
\frac{
|\Delta_\lambda(\vb*{k})|
}{E_\lambda(\vb*{k})}.
\end{split}
\end{equation}
 The diamagnetic contributions from each band $D_z^{(2)\lambda}$ are positive as plotted in Fig.~\ref{Dz2}, indicating that the intraband Cooper pairs of each band flow in the same direction. Furthermore, Figs.~\ref{Dz_1} and \ref{Dz2} reveal that in the absence of the accidental nodal structure of the superconducting gap, the dominant contribution to the electric current $j_z$ originates from the diamagnetic current $j_z^{(2)}$. These interpretations of the supercurrent apply to most superconductors and will be the basis for the following analysis of the spin supercurrent.

\subsection{Spin current}
Next, we present the results for the spin current. The spin current is generated in the second order of Cooper pairs' momentum and is quantified by the spin superfluid density $D_z^{s_z}$. The spin current also shows contrasting behaviors between the weakly and strongly parity-mixed superconducting states. In the $s$$+$$p$-wave state, the spin superfluid density shown in Fig.~\ref{Dzsz}(a) exhibits two peaks around $\psi_s/\psi\simeq 0.5$, with the peak widths narrowing as decreasing temperature. In contrast, in weakly parity-mixed superconductors ($\psi_t/\psi\simeq 0,1$), the spin current remains nearly independent of temperature.

The temperature dependence of the spin superfluid density in the $s$-wave, $d$-wave, and $s$$+$$p$-wave ($\psi_t/\psi = 0.4$) superconducting states is illustrated in Fig.~\ref{Dzsz}(b). The results show that the spin current is almost identical between the $s$-wave and $d$-wave superconductors. However, the $s$$+$$p$-wave state shows a different behavior with a significant temperature dependence of the spin current generation. This feature of the spin current is attributed to the large DOS shown in Fig.~\ref{DOS}.  

To clarify the mechanism of the spin current generation, we perform an analysis similar to that we have done for the electric current. The spin current $j_z^{s_z}$ can be expressed as a sum of three terms, as follows (see the Methods section for derivation).
\begin{equation} 
j_{z}^{s_z}(q) = {\rm const.} + j_{z}^{s_z(1)}(q) + j_{z}^{s_z(2)}(q) + j_{z}^{s_z(3)}(q),
\end{equation} 
\begin{equation} 
\begin{split}
j_{z}^{s_z(1)}(q)
= &\frac{2}{V}\sum_{\vb*{k}} v_{k_z}^{(1)}n_{\vb*{k}}^{(1)}q^2 + \mathcal{O}(q^4)\\
= & D_z^{s_z(1)}q^2 + \mathcal{O}(q^4),
\end{split}
\end{equation}
\begin{equation}
\begin{split}
j_z^{s_z(2)}(q) 
&=\frac{2}{V}\sum_{\vb*{k}} v_{k_z}^{(0)}n_{\vb*{k}}^{(2)}q^2 + \mathcal{O}(q^4)\\
&= D_z^{s_z(2)} q^2 + \mathcal{O}(q^4),
\end{split}
\end{equation} 
\begin{equation} 
\begin{split}
j_z^{s_z(3)} 
&=\frac{2}{V}\sum_{\vb*{k}} v_{k_z}^{(2)}n_{\vb*{k}}^{(0)}q^2 + \mathcal{O}(q^4)\\
&= D_z^{s_z(3)} q^2 + \mathcal{O}(q^4).
\end{split}
\end{equation} 
The spin current $j_z^{s_z(1)}$ is carried by Cooper pairs with center-of-mass velocity $v_{k_z}^{(1)} q$ and spin magnetization $2n_{\vb*{k}}^{(1)}q$. The spin current $j_z^{s_z(2)}$ arises from a modification in the distribution of electrons forming Cooper pairs proportional to the square of the electric current. The spin current $j_z^{s_z(3)}$ is carried by electrons with antiparallel spins, having a relative velocity $v_{k_z}^{(2)}q^2$ and a density $n_{\vb*{k}}^{(0)}$. In the continuum limit, the Hamiltonian in the normal state is given by $H_{\rm N} = \vb*{k}^2/2m -\mu + \alpha \vb*{k}\vdot \vb*{\sigma}$, and thus we obtain $v_{k_z}^{(1)}q = q/2m$ and $v_{k_z}^{(2)}=0$. Consequently, the spin current $j_{z}^{s_z(1)}$ is expressed as a product of the center-of-mass velocity and the bulk spin magnetization, i.e., $j_{z}^{s_z(1)} = S_zq/2m$. Note that $j_z^{s_z(3)}$ vanishes in the continuum limit. 

Let us show the results for our model. The components of the spin superfluid density $D_z^{s_z(1)}$ and $D_z^{s_z(2)}$ in the $s$$+$$p$-wave superconducting states are shown in Figs.~\ref{Dzsz}(c) and \ref{Dzsz}(d), respectively. In weakly parity-mixed superconductors ($\psi_t/\psi \simeq 0, 1$), the spin current $j_z^{s_z(1)}$ dominates the total spin current $j_z^{s_z}$. In contrast, in strongly parity-mixed superconductors ($\psi_t/\psi\simeq 0.5$), the spin currents $j_z^{s_z(1)}$ and $j_z^{s_z(2)}$ are comparable. Although the temperature dependences of $D_z^{s_z(1)}$ and $D_z^{s_z(2)}$ are negligible in weakly parity-mixed superconductors, both of them exhibit temperature dependence in strongly parity-mixed superconducting states. These two components of the spin current determine the total spin current in the whole parameter range.

\begin{figure}[ht]
	\includegraphics[width=2.8in]{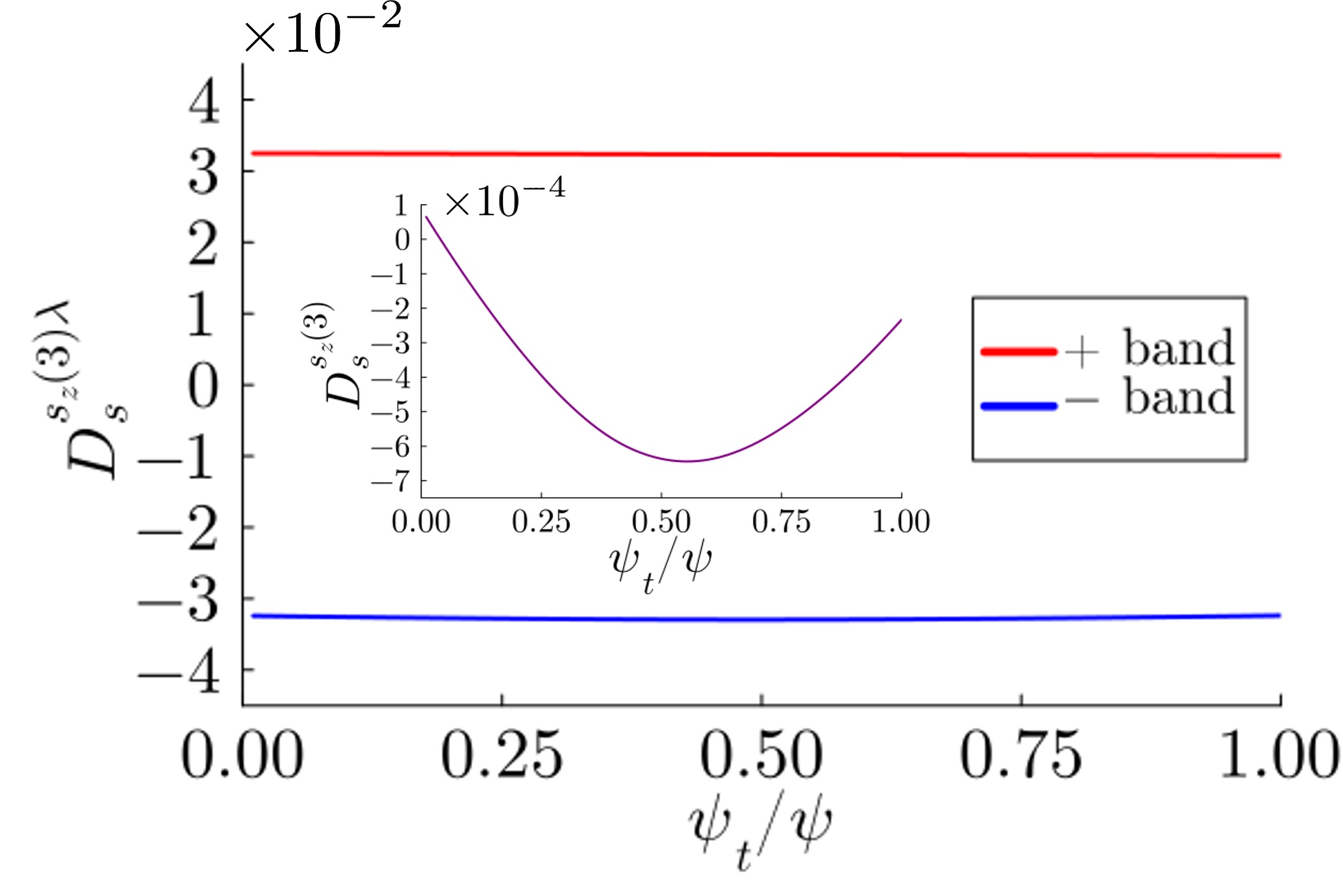}
	\caption{Components of the spin superfluid density $D_z^{s_z(3)\lambda}$ for $s$$+$$p$-wave superconductors at $T/\psi = 0.1$. The inset shows $D_z^{s_z(3)} = D_z^{s_z(3)+} + D_z^{s_z(3)-}$.}
	\label{Dzsz3}
\end{figure}

The remaining component of the spin current $j_z^{s_z(3)}$ can be expressed as a sum of contributions from the two bands (see the Methods section for derivation).
\begin{equation}
\begin{split}
D_z^{s_z(3)} =& D_z^{s_z(3)+} + D_z^{s_z(3)-},\\
D_z^{s_z(3)\lambda}=&-\frac{1}{4V}\sum_{\vb*{k}}(2t\sin k_z + \alpha \cos k_z)\\
&\times\cos \alpha_{\vb*{k}}^\lambda \frac{1 + \lambda\cos\theta_{\vb*{k}}}{2} \qty(f(E_\lambda(\vb*{k}))- \frac{1}{2}).
\end{split}
\end{equation}
Each term in the component $D_z^{s_z(3)}$ is plotted in Fig.~\ref{Dzsz3}. The result shows that the spin current $j_z^{s_z(3)}$ is carried in opposite directions by intraband Cooper pairs in the $\pm$ bands. This result is consistent with the fact that the spins and momentum, and thus velocities, of the electrons forming Cooper pairs are anti-parallel, and the spin orientation of the electron pairs is opposite between bands. Consequently, the spin current $j_z^{s_z(3)}$ is almost canceled between the contributions of the two bands, and therefore, it is negligible compared with the other spin currents $j_z^{s_z(1)}$ and $j_z^{s_z(2)}$, as shown in the inset of Fig.~\ref{Dzsz3}.

\subsection{Current-induced spin current}
\begin{figure}[h]
\includegraphics[width=3.0in]{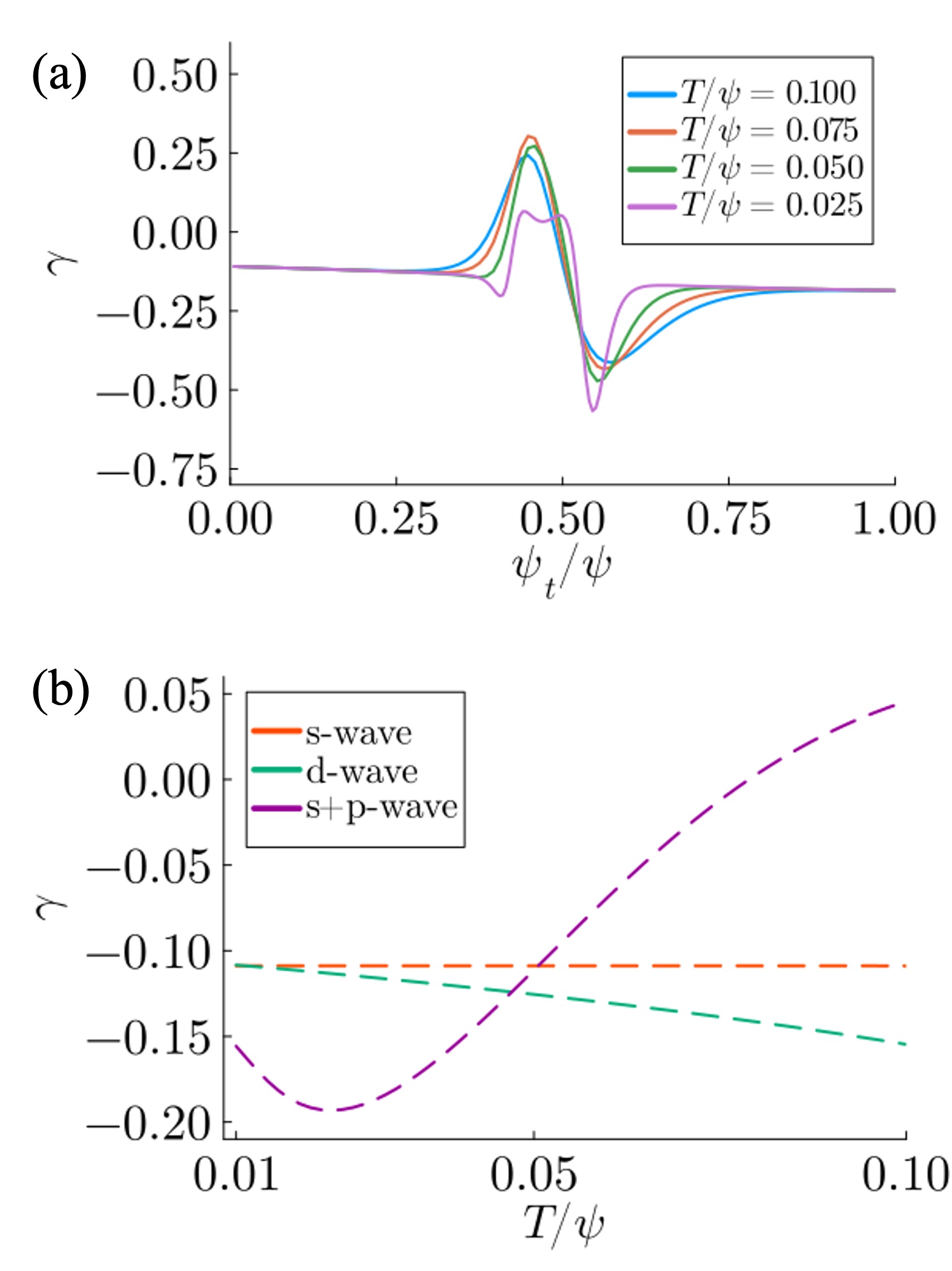}
	\caption{(a) The coefficient $\gamma = D_{z}^{s_z}/(D_z)^{2}$ quantifying the current-induced spin current. We show the order parameter dependence in the $s$$+$$p$-wave superconducting state at various temperatures. (b) Temperature dependence of $\gamma$ in the $s$-wave, $d$-wave and $s$$+$$p$-wave ($\psi_t/\psi = 0.4$) superconducting states.}
	\label{fig:gamma}
\end{figure}

Finally, we show the coefficient $\gamma$ defined with $D_z$ and $D_z^{s_z}$ in Eq.~(\ref{gamma}), which quantifies the spin current under the setup with fixed electric current. The order parameter dependence of $\gamma$ in the $s$$+$$p$-wave superconducting state is presented in Fig.~\ref{fig:gamma}. Notably, the current-induced spin current can be enhanced in strongly parity-mixed superconductors. In contrast, the difference among the $s$-wave, $p$-wave, and $d$-wave states is small. This can be attributed to the sufficiently small DOS at low energies in these superconducting states. Although spin-triplet superconductivity generally possesses spin degrees of freedom, this model assumes a unitary spin-triplet pairing state in spin space. As a result, there is no significant difference between the spin-singlet and spin-triplet states in terms of current-induced spin current, which is a response related to spin. It is an intriguing issue to study spin current in nonunitary spin-triplet superconductors. Another distinguishing feature is that the coefficient $\gamma$ for strongly parity-mixed superconductors exhibits a pronounced temperature dependence, whereas the temperature dependence is negligible in the case of weak parity mixing.

In light of the preceding analyses, we discuss the origin of the spin current generated by the supercurrent. In weakly parity-mixed superconductors, the dominant contribution of the current-induced spin current arises from the spin current $j_z^{s_z(1)}$ produced by the diamagnetic current $j_z^{(2)}$. On the other hand, in superconductors with strong parity mixing which gives rise to gap nodes or gap anisotropy, the
primary contributions come from the spin current $j_z^{s_z(1)}$ and $j_z^{s_z(2)}$ produced by the paramagnetic current $j_z^{(1)}$ and the diamagnetic current $j_z^{(2)}$.

\section{Discussion}
In this paper, we have investigated the nonreciprocal spin current produced by the electric current in chiral-structured superconductors with various order parameters. 
We have classified the spin current and demonstrated that the origin of the current-induced spin current depends on the superconducting order parameter, in particular the magnitude of parity mixing. Due to space-inversion symmetry breaking, the superconducting order parameters for the even-parity spin-singlet pairing and the odd-parity spin-triplet pairing are hybridized. In weakly parity-mixed superconductors, the spin current is almost temperature-independent and arises from spin-polarized Cooper pairs with finite center-of-mass momentum. The spin current of this origin constitutes the dominant contribution to the CISS effect. On the other hand, in strongly parity-mixed superconductors where a band-dependent or even nodal superconducting gap is realized, the temperature-dependent spin current is produced and is also carried by electrons with opposite momentum and antiparallel spins that form Cooper pairs. Furthermore, the CISS effect can be strongly enhanced at low temperatures because of the strong parity mixing, which is attributed to the quasiparticle excitation. 
While we have studied a two-dimensional model, essentially the same features are expected to appear in three-dimensional isotropic superconductors with structural chirality where the $\vb*{g}$-vector of ASOC and the spin texture in the momentum space are aligned in a similar way to the two-dimensional systems. This study may pave a route for an efficient CISS effect in superconductors. Moreover, the electric current in superconductors does not generate Joule heating, which presents a challenge for spintronics applications, and thus the spin current induced by supercurrent could be a crucial step toward realizing non-dissipative spintronics~\cite{Grein2009-za,Alidoust2010-tr,Yang2010-lv,Eschrig2011-lp,Bergeret2014-lw,Linder2015-kt,Konschelle2015-rw,Jacobsen2016-pu,Linder2017-nh,He2019-hz}.

As a remark, broken inversion symmetry generally suppresses the interband pairing and thus a part of the spin-triplet pairing.
For this reason, we have assumed intraband pairing states, which is justified when the band splitting in the normal state significantly exceeds the superconducting energy scale, i.e. $\alpha \gg T_{\rm c}$~\cite{Bauer2012-ix}.  The interband pairing may play a role when the spin-orbit coupling is comparable to the superconducting gap, and the study of interband pairing will be an interesting future issue. 

A large spin-orbit coupling is a desirable feature for the strong parity mixing in superconducting order parameters. A chiral-structured material with a large ASOC is recently discovered superconductors \ce{IrGe4} and \ce{RhGe4}~\cite{Nakamura-jpsj-2023}, where
DOS on the Fermi surfaces of each band can be 
different due to the chiral ASOC. Furthermore, a recent experiment has reported a large effective ASOC in a chiral organic superconductor, leading to a singlet-triplet mixed superconducting state~\cite{sato2025}.

Another important issue is the mechanism of spin accumulation at the ends of the superconductor. In superconductors, the direct electric current can flow in equilibrium without dissipation, and thus time-reversal symmetry breaking due to nonequilibrium dynamics can be avoided. Consequently, the time-reversal parity of physical quantities gives constraints on the response to external fields. Spin magnetization has the same time-reversal parity as the electric current, while it is opposite to the spin current. Therefore, it is concluded that the spin current does not lead to spin accumulation when only the supercurrent is present. However, the dissipative electric current which breaks time-reversal symmetry in the nonequilibrium sense can penetrate into superconductors over a length scale comparable to or larger than the coherence length~\cite{kopnin2001,Artemenko_1979,kopnin1984}. This normal current causes dissipation at the ends of the superconductors, and the resulting nonequilibrium state can host antiparallel spin accumulation. Therefore, the CISS effect could be observed in chiral-structured superconductors under a direct electric current. Estimation of the spin accumulation is desirable but beyond the scope of this paper.

\begin{figure}[t]
\includegraphics[width=2.6in]{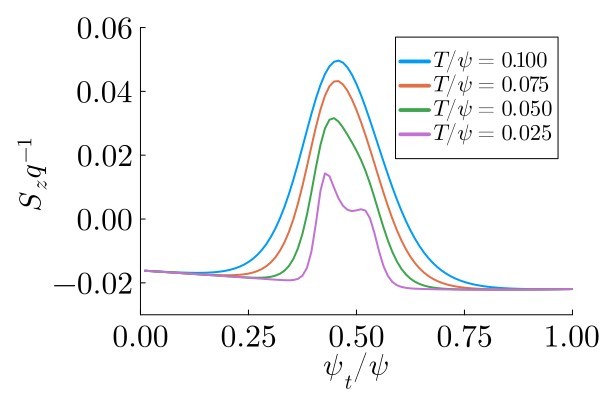}
\caption{The order parameter dependence of the Edelstein coefficient $S_z q^{-1}=V^{-1}\sum_{\vb*{k}}2n_{\vb*{k}}^{(1)}$ for $s$$+$$p$-wave superconductors.}
\label{spin}
\end{figure}

Finally, we briefly discuss current-induced spin magnetization in bulk superconductors, known as the Edelstein effect. This phenomenon can be distinguished from the CISS effect because the spin magnetization occurs not only at the ends of the superconductor but also in the bulk region. Furthermore, the magnetization due to the Edelstein effect is expected to be parallel between the ends, in contrast to the CISS effect. Figure~\ref{spin} shows the order parameter dependence of the bulk Edelstein coefficient $S_z/q = \sum_{\vb*{k}}2n_{\vb*{k}}^{(1)}/V$ for $s$$+$$p$-wave superconductors at various temperatures. The results indicate that the admixture of the spin-singlet pairing and the spin-triplet pairing enables quasiparticle excitation and leads to the enhancement and sign reversal of spin magnetization due to the Edelstein effect. It is worth noting that, in weakly parity-mixed superconductors, the spin current $j_z^{s_z(1)}\simeq S_zq/2m$ dominates the total spin current, where $S_z$ represents spin magnetization in the bulk superconductor. In this case, the spin accumulation at the ends of the superconductor arising from the spin current is considered to have the same origin as the bulk spin magnetization.

\section*{Acknowledgements}

The authors acknowledge useful discussions with Yuma Hirobe, Shun Asano, Shin Kaneshiro, Ryotaro Sano, Daichi Hirobe, Atsuo Shitade, and Hiroshi Yamamoto.
This work was supported by JSPS KAKENHI (Grant Numbers JP22H01181, JP22H04933, JP23K17353, JP23K22452, JP24K21530, JP24H00007).

\section{Methods}
\subsection{Formulation of electric current and spin current}
The electric current operator and the spin current operator are defined as
\begin{equation}
j_z
=\frac{1}{V}\sum_{\vb*{k}ss'}
\pdv{H_{\rm N}(\vb*{k})_{ss'}}{k_z}
c_{\vb*{k}s}^\dag c_{\vb*{k}s'},
\end{equation} 
\begin{equation} 
j_{z}^{s_z}
=\frac{1}{2}\{\hat{S}_z, \hat{j}_z\}.
\end{equation} 
The expectation values of electric current and electric current are given by
\begin{equation}
\begin{split}
j_z(q) 
=&\frac{1}{V}\sum_{\vb*{k}}v_{k_z+q/2\uparrow}  \braket{n_{\vb*{k}+q/2\uparrow}}_q\\
&+v_{-k_z+q/2\downarrow} \braket{n_{-\vb*{k}+q/2\downarrow}}_q,
\end{split}
\end{equation} 
\begin{equation} 
\begin{split}
j_{z}^{s_z}(q) 
=&\frac{1}{V}\sum_{\vb*{k}}v_{k_z+q/2\uparrow}  \braket{n_{\vb*{k}+q/2\uparrow}}_q\\
&-v_{-k_z+q/2\downarrow} \braket{n_{-\vb*{k}+q/2\downarrow}}_q,
\end{split}
\end{equation} 
where $\hat{S}_z =\sum_{\vb*{k}ss}\sigma_z c_{\vb*{k}s}^\dag c_{\vb*{k}s'}$. To evaluate the expectation values, 
we introduce a Bogoliubov-de Gennes (BdG) Hamiltonian $H(\vb*{k},q)$ by
\begin{equation}
H_{\rm MF}^{q}
=
\frac{1}{2}\sum_{\vb*{k}}
\psi^\dag(\vb*{k},q)H(\vb*{k},q){\psi}(\vb*{k},q) + {\rm const.},
\end{equation}
\begin{equation} 
H(\vb*{k},q) 
= \mqty(H_{\rm N}(\vb*{k}+\frac{q}{2})&\Delta(\vb*{k})\\ \Delta^\dag(\vb*{k})&-H_{\rm N}^T(-\vb*{k}+\frac{q}{2})),
\end{equation} 
where
$ {\psi} (\vb*{k},q) 
= (c_{\vb*{k}+\frac{q}{2}\uparrow},c_{\vb*{k}+\frac{q}{2}\downarrow},
c_{-\vb*{k}+\frac{q}{2}\uparrow}^\dag,c_{-\vb*{k}+\frac{q}{2}\downarrow}^\dag)^{\rm T}$ is the Nambu spinor.
After a unitary transformation, the BdG Hamiltonian turned into a diagonal matrix
\begin{equation} 
\begin{split}
& U^\dag(\vb*{k},q)H(\vb*{k},q) U(\vb*{k},q)\\
=& {\rm diag}(E_1(\vb*{k},q),E_2(\vb*{k},q),E_3(\vb*{k},q),E_4(\vb*{k},q)).
\end{split}
\end{equation} 
Here, $E_i(\vb*{k},q)\ (1\leq i \leq 4)$ are the energy eigenvalues of Bogoliubov quasiparticles, and the operators for Bogoliubov quasiparticles are given by $\Phi(\vb*{k},q)
= U^\dag (\vb*{k},q){\psi} (\vb*{k},q)$. We obtain expectation values of the electric current and the spin current as follows, 
\begin{equation} 
\begin{split}
j_z(q) =\frac{1}{V}\sum_{i\vb*{k}}v_{k_z+q/2\uparrow}  |U_{1i}(\vb*{k},q)|^2f(E_i(\vb*{k},q))&\\
+v_{-k_z+q/2\downarrow}|U_{4i}(\vb*{k},q)|^2f(-E_i(\vb*{k},q))&,
\end{split}
\end{equation} 
\begin{equation} 
\begin{split}
j_z^{s_z}(q) =\frac{1}{V}\sum_{i\vb*{k}}v_{k_z+q/2\uparrow}  |U_{1i}(\vb*{k},q)|^2f(E_i(\vb*{k},q))&\\
-v_{-k_z+q/2\downarrow}|U_{4i}(\vb*{k},q)|^2f(-E_i(\vb*{k},q))&,
\end{split}
\end{equation} 
with the Fermi distribution function $f(x)$. 

To clarify the origins of electric current and spin current, we divide the expectation values 
into some terms. First, the relation $\braket{n_{\vb*{k}+q/2\uparrow}}_q = \braket{n_{-\vb*{k}-q/2\downarrow}}_{-q}$ is obtained due to the time-reversal symmetry. Thus, the expectation value for electron density is written as 
\begin{equation} 
\braket{n_{\vb*{k}+q/2\uparrow}}_q 
= n_{\vb*{k}}^{(0)}+ n_{\vb*{k}}^{(1)} q + n_{\vb*{k}}^{(2)} q^2
+ \mathcal{O}(q^3),
\end{equation} 
\begin{equation} 
\braket{n_{-\vb*{k}+q/2\downarrow}}_q
= n_{\vb*{k}}^{(0)} - n_{\vb*{k}}^{(1)} q + n_{\vb*{k}}^{(2)} q^2
+ \mathcal{O}(q^3).
\end{equation}
Furthermore, $v_{k_z+q/2\uparrow} = -v_{-k_z-q/2\downarrow}$ holds because of the time-reversal symmetry. Thus, the velocity $v_{k_zs}$ is given as follows,
\begin{equation} 
v_{k_z+q/2\uparrow}
= v_{k_z}^{(0)} + v_{k_z}^{(1)}q +v_{k_z}^{(2)}q^2
+ \mathcal{O}(q^3),
\end{equation} 
\begin{equation} 
v_{-k_z+q/2\downarrow}
= -v_{k_z}^{(0)} + v_{k_z}^{(1)}q -v_{k_z}^{(2)}q^2
+ \mathcal{O}(q^3).
\end{equation} 
Using the above relations, we obtain the paramagnetic current $j_z^{(1)}$ and the diamagnetic current $j_z^{(2)}$ as
\begin{equation} 
j_z(q) = j_z^{(1)}(q) + j_z^{(2)}(q),
\end{equation} 
\begin{equation} 
\begin{split}
j_z^{(1)}(q)
&= \frac{2}{V}\sum_{\vb*{k}}v_{k_z}^{(0)}n_{\vb*{k}}^{(1)}q + \mathcal{O}(q^2)\\
&= D_z^{(1)} q + \mathcal{O}(q^2),
\end{split}
\end{equation} 
\begin{equation} 
\begin{split}
j_z^{(2)}(q)
&= \frac{2}{V}\sum_{\vb*{k}}v_{k_z}^{(1)}n_{\vb*{k}}^{(0)}q + \mathcal{O}(q^2)\\
&= D_z^{(2)} q + \mathcal{O}(q^2).
\end{split}
\end{equation} 
In the same way the spin current consists of three terms as follows.
\begin{equation} 
j_{z}^{s_z}(q)=
{\rm const.} + j_{z}^{s_z(1)}(q) + j_{z}^{s_z(2)}(q) + j_{z}^{s_z(3)}(q),
\end{equation} 
\begin{equation} 
\begin{split}
j_{z}^{s_z(1)}(q)
=&\frac{2}{V}\sum_{\vb*{k}}
v_{k_z}^{(1)}n_{\vb*{k}}^{(1)}q^2 + \mathcal{O}(q^4)\\
= & D_z^{s_z}q^2 + \mathcal{O}(q^4),
\end{split}
\end{equation}
\begin{equation}
\begin{split}
j_z^{s_z(2)}(q) 
&=\frac{2}{V}\sum_{\vb*{k}}
v_{k_z}^{(0)}n_{\vb*{k}}^{(2)}q^2 + \mathcal{O}(q^4)\\
&= D_z^{s_z(2)} q^2 + \mathcal{O}(q^4),
\end{split}
\end{equation} 
\begin{equation} 
\begin{split}
j_z^{s_z(3)} &=
\frac{2}{V}\sum_{\vb*{k}} 
v_{k_z}^{(2)}n_{\vb*{k}}^{(0)}q^2 + \mathcal{O}(q^4)\\
&= D_z^{s_z(3)} q^2 + \mathcal{O}(q^4).
\end{split}
\end{equation} 
\subsection{Expectation value of electron density}
The eigenvalue equation of the BdG Hamiltonian in the absence of the electric current is given by $H(\vb*{k},0) u_i^{(0)}(\vb*{k}) =  E_i(\vb*{k})u_i^{(0)}(\vb*{k})\ (1\leq i \leq 4)$. The eigen values and eigen vectors of $H(\vb*{k},0)$ are obtained as
\begin{equation} 
\begin{split}
\label{eigenval}
 E_1(\vb*{k}) &=  - E_2(\vb*{k}) = E_+(\vb*{k}),\\
E_3(\vb*{k}) &= - E_4(\vb*{k}) = E_-(\vb*{k}),
\end{split}
\end{equation} 
\begin{equation} 
u_1^{(0)}(\vb*{k})
= \mqty(
\cos\frac{\theta_{\vb*{k}}}{2}\cos\frac{\alpha_{\vb*{k}}^{+}}{2}\\
\sin\frac{\theta_{\vb*{k}}}{2}\cos\frac{\alpha_{\vb*{k}}^{+}}{2}\\
\sin\frac{\theta_{\vb*{k}}}{2}\sin\frac{\alpha_{\vb*{k}}^{+}}{2}\\
-\cos\frac{\theta_{\vb*{k}}}{2}\sin\frac{\alpha_{\vb*{k}}^{+}}{2}
),
\end{equation} 
\begin{equation} 
u_2^{(0)}(\vb*{k})
= \mqty(
-\cos\frac{\theta_{\vb*{k}}}{2}\sin\frac{\alpha_{\vb*{k}}^{+}}{2}\\
-\sin\frac{\theta_{\vb*{k}}}{2}\sin\frac{\alpha_{\vb*{k}}^{+}}{2}\\
\sin\frac{\theta_{\vb*{k}}}{2}\cos\frac{\alpha_{\vb*{k}}^{+}}{2}
\\
-\cos\frac{\theta_{\vb*{k}}}{2}\cos\frac{\alpha_{\vb*{k}}^{+}}{2}),
\end{equation}
\begin{equation} 
u_3^{(0)}(\vb*{k})
= \mqty(
-\sin\frac{\theta_{\vb*{k}}}{2}\cos\frac{\alpha_{\vb*{k}}^{-}}{2}\\
\cos\frac{\theta_{\vb*{k}}}{2}\cos\frac{\alpha_{\vb*{k}}^{-}}{2}\\
\cos\frac{\theta_{\vb*{k}}}{2}\sin\frac{\alpha_{\vb*{k}}^{-}}{2}
\\
\sin\frac{\theta_{\vb*{k}}}{2}\sin\frac{\alpha_{\vb*{k}}^{-}}{2}
),
\end{equation} 
\begin{equation} 
u_4^{(0)}(\vb*{k})
= \mqty(
\sin\frac{\theta_{\vb*{k}}}{2}\sin\frac{\alpha_{\vb*{k}}^{-}}{2}\\
-\cos\frac{\theta_{\vb*{k}}}{2}\sin\frac{\alpha_{\vb*{k}}^{-}}{2}\\
\cos\frac{\theta_{\vb*{k}}}{2}\cos\frac{\alpha_{\vb*{k}}^{-}}{2}
\\
\sin\frac{\theta_{\vb*{k}}}{2}\cos\frac{\alpha_{\vb*{k}}^{-}}{2}
),
\end{equation} 
where $\theta_{\vb*{k}}$ and $\alpha_{\vb*{k}}^\lambda$ are defined in Eq.~(\ref{alpha_def}). Consequently, the expectation value of the electron density in the absence of the electric current is written as 
\begin{equation} 
\begin{split}
n_{\vb*{k}}^{(0)}
= &
 \frac{1}{2} 
 +  \cos \alpha_{\vb*{k}}^+ 
 \frac{
 1 + \cos\theta_{\vb*{k}}
 }{2} 
 \qty( 
 f(E_+(\vb*{k}))- \frac{1}{2}
 )\\
 &   
 + \cos \alpha_{\vb*{k}}^- 
 \frac{
 1 - \cos\theta_{\vb*{k}}
 }{2} 
 \qty( 
 f(E_-(\vb*{k}))- \frac{1}{2}
 ).
\end{split}
\end{equation} 

\newpage
\bibliography{ref.bib}

\end{document}